# MetricGAN+: An Improved Version of MetricGAN for Speech Enhancement

*Szu-Wei Fu[1], Cheng Yu[1], Tsun-An Hsieh[1], Peter Plantinga[2], Mirco Ravanelli[3], Xugang Lu[4], Yu Tsao[1]*

[1] Research Center for Information Technology Innovation, Academia Sinica, Taipei, Taiwan
[2] The Ohio State University, Columbus, OH, USA
[3] Mila-Quebec AI Institute, Montreal, Canada
[4] National Institute of Information and Communications Technology, Kyoto, Japan

{jasonfu,yu.tsao}@citi.sinica.edu.tw

## Abstract

The discrepancy between the cost function used for training a speech enhancement model and human auditory perception usually makes the quality of enhanced speech unsatisfactory. Objective evaluation metrics which consider human perception can hence serve as a bridge to reduce the gap. Our previously proposed MetricGAN was designed to optimize objective metrics by connecting the metric with a discriminator. Because only the scores of the target evaluation functions are needed during training, the metrics can even be non-differentiable. In this study, we propose a MetricGAN+ in which three training techniques incorporating domain-knowledge of speech processing are proposed. With these techniques, experimental results on the VoiceBank-DEMAND dataset show that MetricGAN+ can increase PESQ score by 0.3 compared to the previous MetricGAN and achieve state-of-the-art results (PESQ score = 3.15).

## 1. Introduction

There are many different applications and goals for a speech enhancement (SE) model. For example, in human-to-human communication, we care about speech quality or intelligibility (e.g., during a phone call with severe background noise, the intelligibility may be more important than quality). On the other hand, in human-to-machine communication, the goal of SE is to improve the speech recognition performance (e.g., reducing the word error rate (WER) under noisy conditions for an automated speech recognition (ASR) system). Therefore, training a task-specific SE model may obtain better performance for its targeted applications.

To deploy a task-specific SE model, the most intuitive way is to adopt a loss function that is relevant to the final goal. Although directly applying a measure based on the difference in signal level (e.g., $L_1$ or $L_2$ loss) is straightforward, several studies have shown that it is not highly correlated to speech quality [1-3], intelligibility [4], and WER [5, 6].

An alternative would be to directly optimize speech quality or intelligibility. This is often very challenging and normally objective evaluation metrics are used as surrogates. Among the human perception-related objective metrics, the perceptual evaluation of speech quality (PESQ) [7] and short-time objective intelligibility (STOI) [8] are two popular functions used to evaluate speech quality and intelligibility, respectively. The design of these two metrics considers human auditory perception and has shown higher correlation to subjective listening tests than simple $L_1$ or $L_2$ distance between clean and degraded speech signals [1, 4].

The current techniques to optimize these objective scores can be categorized into two types depending on whether the details of evaluation metrics have to be known: 1) white-box: these methods [4, 9-12] approximate the complex evaluation metrics with a hand-crafted, differentiable one. However, the details of the metrics have to be known and it can only be used for the targeted metric. (2) black-box: these methods [3, 13, 14] mainly treat the metric as a reward and apply reinforcement learning based techniques to increase the scores. However, the training is usually inefficient with limited performance improvement.

MetricGAN [15] falls into the black-box category, and it can achieve better training efficiency and moderate performance improvement (the average PESQ score increases more than 0.1) compared to conventional $L_1$ loss. Although MetricGAN can be easily applied to optimize different evaluation metrics (e.g., PESQ, STOI, or WER), we mainly considered PESQ score optimization as an example. Other extensions can be found at [16-18].

In this study, to further boost the performance of the MetricGAN framework and reveal the important factors that affect the performance, we propose MetricGAN+. The basic idea behind MetricGAN+ does not change and the improvement comes from including three training techniques that incorporate domain-knowledge of speech processing. Two improvements are proposed for the discriminator (*D*) and one for the generator (*G*):

For the discriminator:

1) ***Include noisy speech for discriminator training***: In addition to enhanced and clean speech, noisy speech is used to minimize the distance between the discriminator and target objective metrics.

2) ***Increase sample size from replay buffer***: Speech generated from the previous epochs is reused for training *D*. This can prevent *D* from catastrophic forgetting [19].

For the generator:

1) ***Learnable sigmoid function for mask estimation***: A conventional sigmoid is not optimal for mask estimation because it is the same for all frequency bands and has a maximum value of 1. A per-frequency learnable sigmoid function is more flexible and improves the performance of SE.

To foster reproducibility, the MetricGAN+ is available within the SpeechBrain toolkit[1].

---
[1] https://speechbrain.github.io/

## 2. Introduction to MetricGAN

The main idea of MetricGAN is to mimic the behavior of a target evaluation function (e.g., PESQ function) with a neural network (e.g., Quality-Net [20]). The surrogate evaluation function is learned from raw scores, treating the target evaluation function as a black box. Once the surrogate evaluation is trained, it can be used as a loss function for the speech enhancement model. Unfortunately, a static surrogate is easily fooled by adversarial examples [22] (estimated quality scores increase but true scores decrease [21]). To mitigate this issue, we recently proposed a learning framework where the surrogate loss and enhancement model are alternately updated [15]. This method is called MetricGAN because its goal is to optimize black-box metric scores, with a training flow that is similar to the one of generative adversarial networks (GANs). Below, we briefly introduce the training of MetricGAN.

Let $Q'(I)$ be a function that represents the target evaluation metric normalized between 0 and 1, where $I$ denotes the input of the metric. For example, for PESQ and STOI, $I$ denotes a pair of enhanced speech, $G(x)$ (or noisy speech, $x$) that we want to evaluate, and its corresponding clean speech, $y$. To ensure that the discriminator network ($D$) behaves similar to $Q'$, the objective function of $D$ is

$$L_{D(\text{MetricGAN})} = \mathbb{E}_{x,y}[(D(y,y) - Q'(y,y))^2 + (D(G(x),y) - Q'(G(x),y))^2] \quad (1)$$

The two terms are used to minimize the difference between $D(.)$ and $Q'(.)$ for clean and enhanced speech, respectively. Note that, $Q'(y,y) = 1$ and $0 \leq Q'(G(x),y) \leq 1$.

The training of the generator network ($G$) can completely rely on the adversarial loss

$$L_{G(\text{MetricGAN})} = \mathbb{E}_x[(D(G(x),y) - s)^2] \quad (2)$$

where $s$ denotes the desired assigned score. For example, to generate clean speech, we can simply assign $s$ to be 1. The overall training flow is shown in Figure 1.

## 3. From MetricGAN to MetricGAN+

To improve the performance of the MetricGAN framework, some advanced learning techniques are proposed. During the investigation, we also study the factors that significantly influence the performance or training efficiency. The improvement of MetricGAN+ mainly comes from the following three modifications.

### 3.1. Learning the Metrics Scores for Noisy Speech

Kawanaka *et al.* [16] proposed to include noisy speech when training the discriminator. This turned out to stabilize the learning process. We adopt the same strategy for MetricGAN+ as well. The loss function of the discriminator network is thus modified as follows:

$$L_{D(\text{MetricGAN+})} = \mathbb{E}_{x,y}[(D(y,y) - Q'(y,y))^2 + (D(G(x),y) - Q'(G(x),y))^2 + (D(x,y) - Q'(x,y))^2] \quad (3)$$

where the third term is used to minimize the difference between $D(.)$ and $Q'(.)$ for noisy speech.

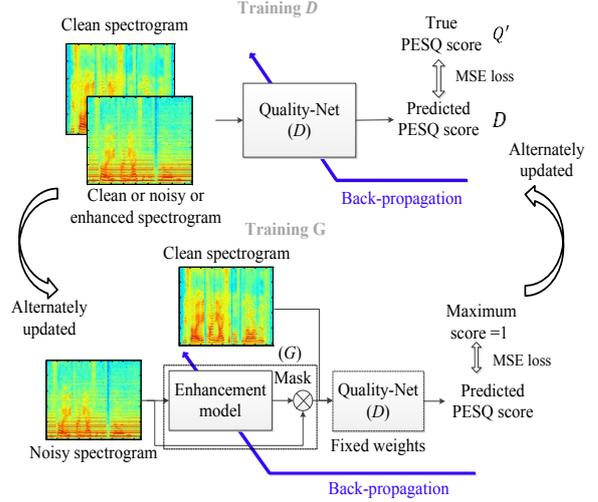

Figure 1: *Training flow of MetricGAN.*

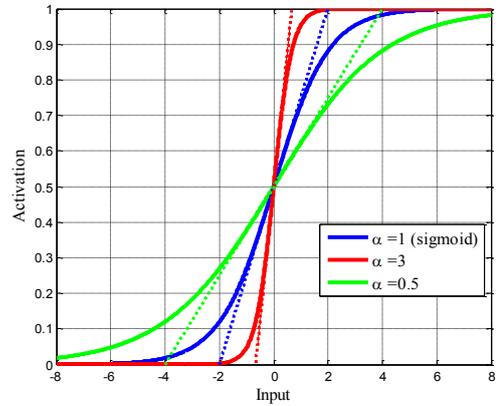

Figure 2: *Sigmoid function with different α.*

### 3.2. Samples from Experience Replay Buffer

As in the deep Q-network [23], we found that reusing the data generated from the previous epochs to train the discriminator brings a huge improvement in performance. Intuitively, without experience replay, the discriminator may forget the behavior of target $Q'$ function at the previous generated speech, thus making $D(.)$ less similar to $Q'(.)$. To illustrate how the replay buffer works, we present the training process in Algorithm 1. In MetricGAN+, we increase **history_portion** from 0.1 (used in MetricGAN) to 0.2.

### 3.3. Learnable Sigmoid Function for Mask Estimation

Most mask-based speech enhancement methods [24] apply a sigmoid activation to the output layer to constrain the mask to be between 0 and 1. However, due to the phase difference, the sum of clean ($|Y(t,f)|$) and noise ($|N(t,f)|$) magnitude spectrograms do not exactly match the noisy ($|X(t,f)|$) magnitude spectrogram [25]:

$$|X(t,f)| \neq |Y(t,f)| + |N(t,f)| \quad (4)$$

The ideal magnitude mask $|Y(t,f)|/|X(t,f)|$ is hence not guaranteed to be smaller than 1. Therefore, we set the scale variable β equal to 1.2 in (5).

**Algorithm 1** Training with replay buffer

for **epoch** in [1, 2,...**Epoch**]**:**
    Randomly sample **number_of_samples** from training Set.
    # *Train G*enerator:
      Use these samples $x$ to train $G$ using (2).
    # *Store into the buffer*:
      Store these samples and the corresponding true scores ($G(x)$, $Q'(G(x), y)$) into the replay buffer.
    # *Train Discriminator*:
    1) Update $D$ with current sampled data $x$ using (3).
    2) Randomly sample **history_portion** of the buffer and update $D$ with the historical enhanced data.
end for

In addition, the standard sigmoid function to compress the input value may not be optimal for speech processing. For example, because the patterns of both noise and speech in high and low frequency bands are distinct, different frequency bands could have their own compression function for mask estimation. To give this flexibility to our model, we design a learnable sigmoid function as follows:

$$y = \beta/(1 + e^{-\alpha x}) \quad (5)$$

where $\alpha$ is learned from training data. Different frequency bands have their own $\alpha$.

$\alpha$ controls the shape of the compression function. As shown in Figure 2, large $\alpha$ (red) behaves like a hard threshold and most output values are either 0 or 1 (more non-smooth and more saturated [26], like a binary mask). On the other hand, small $\alpha$ (green) behaves more like a linear function, which can be observed that it has larger overlap with the green dotted line (the overlap between the dotted lines and sigmoid functions can roughly show the range of linearity).

## 4. Experiments

### 4.1. Dataset

To compare the proposed MetricGAN+ with other existing methods, we use the publicly available VoiceBank-DEMAND dataset [27]. The train sets (11572 utterances) consists of 28 speakers with 4 signal-to-noise ratio (SNR) (15, 10, 5, and 0 dB). The test sets (824 utterances) consists of 2 speakers with 4 SNR (17.5, 12.5, 7.5, and 2.5 dB). Details about the data can be found in the original paper. In addition to the PESQ score, we evaluate the performance with other three metrics: CSIG predicts the mean opinion score (MOS) of the signal distortion, CBAK predicts the MOS of the background noise interferences, and COVL predicts the MOS of the overall speech quality. All these three metrics range from 1 to 5. For all metrics used, higher values indicate better performance.

### 4.2. Model Structure

The generator used in this experiment is a BLSTM [28] with two bidirectional LSTM layers, with 200 neurons each. The LSTM is followed by two fully connected layers, each with 300 LeakyReLU nodes and 257 (learnable) sigmoid nodes for mask estimation, respectively. When this mask is multiplied with the input noisy magnitude spectrogram, the noise components should be removed. In addition, as reported in [3], to prevent musical noise, flooring was applied to the estimated mask before T-F mask processing. Here, we set the lower threshold of the T-F mask to 0.05. The discriminator herein is a CNN with four two dimensional (2-D) convolutional layers with 15 filters and a kernel size of (5, 5). To handle the variable length input (different speech utterances have different length), a 2-D global average pooling layer was added such that the features can be fixed at 15 dimensions (15 is the number of feature maps in the previous layer). Three fully connected layers were added subsequently, each with 50 and 10 LeakyReLU neurons, and 1 linear node. In addition, to make $D$ a smooth function (ie., no small modification in the input spectrogram can make a significant change to the estimated score), the discriminator is constrained to be 1-Lipschitz continuous by using spectral normalization [29]. **number_of_samples** is set to 100 in the experiment.

Table 1: *Ablation study of MetricGAN+.*

| | PESQ | CSIG | CBAK | COVL |
|---|---|---|---|---|
| MetricGAN [15] | 2.86 | 3.99 | 3.18 | 3.42 |
| - Input normalization | 2.95 | 4.03 | 3.11 | 3.49 |
| + Include noisy (Sec. 3.1) | 3.02 | 4.13 | **3.23** | 3.57 |
| + Increase **history_portion** from replay buffer (Sec. 3.2) | 3.05 | 4.11 | 3.15 | 3.57 |
| + Learnable sigmoid (Sec. 3.3) = MetricGAN+ | **3.15** | **4.14** | 3.16 | **3.64** |

Table 2: *Results of different **history_portions**.*

| history_portion | PESQ | CSIG | CBAK | COVL |
|---|---|---|---|---|
| 0 | 2.82 | 3.99 | 3.38 | 3.41 |
| 0.1 | 3.02 | 4.13 | 3.23 | 3.57 |
| 0.2 | 3.05 | 4.11 | 3.15 | 3.57 |

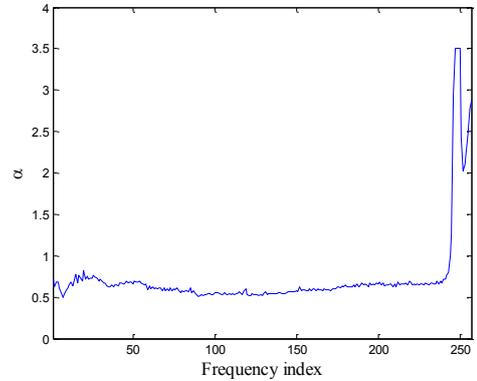

Figure 3: *Learned values of $\alpha$ in learnable sigmoid function.*

### 4.3. Experimental Results

We first show the effects of the different training techniques introduced in Section 3. In Table 1, the results in each row are achieved with the setting from the previous row plus the current technique. From the table, it emerges that removing the input spectrogram normalization and including noisy speech for $D$ training leads to a larger improvement than further increasing the sample size from the buffer. However, Table 2 shows that if we do not keep the replay buffer, the PESQ score only reaches 2.82. When we randomly sample 10% and 20% of historical enhanced data from the buffer, the score increases by 0.2 and 0.23, respectively. We do not observe further improvement with ***history_portion***=0.3. This implies that without the buffer, the discriminator may just focus on the evaluation results on the current samples and ignore its correctness on the previously generated speech (catastrophic forgetting [19]). Due to the discrepancy between $D$ and $Q'$, the gradient of $D$ may thus not a good approximation for that of $Q'$.

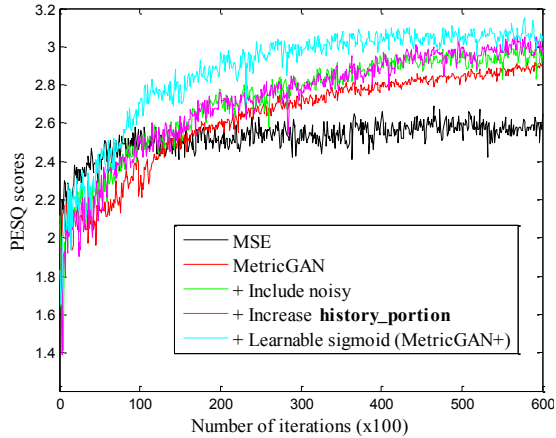

Figure 4: *Learning curves of different settings (structure of G is fixed).*

Table 3: *Compared MetricGAN+ with other methods on the VoiceBank-DEMAND dataset.*

|  | PESQ | CSIG | CBAK | COVL |
|---|---|---|---|---|
| Noisy | 1.97 | 3.35 | 2.44 | 2.63 |
| SEGAN [36] | 2.16 | 3.48 | 2.94 | 2.80 |
| MMSE-GAN [37] | 2.53 | 3.80 | 3.12 | 3.14 |
| SERGAN [38] | 2.62 | - | - | - |
| BLSTM (MSE) | 2.71 | 3.94 | 3.28 | 3.32 |
| MetricGAN [15] | 2.86 | 3.99 | 3.18 | 3.42 |
| HiFi-GAN [32] | 2.94 | 4.07 | 3.07 | 3.49 |
| DeepMMSE [39] | 2.95 | 4.28 | 3.46 | 3.64 |
| MHSA+SPK [30] | 2.99 | 4.15 | 3.42 | 3.57 |
| PHASEN [34] | 2.99 | 4.21 | 3.55 | 3.62 |
| SDR-PESQ [11] | 3.01 | 4.09 | 3.54 | 3.55 |
| T-GSA [31] | 3.06 | 4.18 | **3.59** | 3.62 |
| DEMUCS [33] | 3.07 | **4.31** | 3.40 | 3.63 |
| **MetricGAN+** | **3.15** | 4.14 | 3.16 | **3.64** |

Table 1 also reveals that applying learnable sigmoid can further increase the scores. We tried to make β learnable as well, but the performance did not further improve. The learned values of $\alpha$ are shown in Figure 3. Most of $\alpha$ are smaller than 1 (original value in the conventional sigmoid function) and close to 0.5. As pointed out in Section 3.3, this implies that for most frequency bins, the learnable sigmoid is behaving more like a linear function. This naturally leads to a gradient back-propagation that is more efficient than the one happening in saturated regions. On the other hand, for high frequency bins, the learned $\alpha$ are much larger than 1, and hence the mask is behaving more like a binary mask. We conjecture this is because the noise types in the training data do not occupy the high frequency regions, or this is due to the characteristics of PESQ function. However, more experiments are needed to verify the possible reasons.

In Figure 4, we show the training curves of MetricGAN with different training techniques and the same BLSTM model structure as the generator in MetricGAN, but trained with MSE loss. From this figure, we can first observe that all MetricGAN-based methods outperform the MSE loss by a large margin. In addition, both including noisy speech for discriminator training and applying a learnable sigmoid not only improve the final performance but also lead to a higher training efficiency.

Table 3 compares the proposed MetricGAN+ with other popular methods. Although our generator is just a conventional BLSTM with magnitude spectrogram as inputs, with appropriate loss function, it outperforms recent models (e.g., attention mechanism [30, 31]) or phase-aware inputs (e.g., waveform [32] [33] or phase [34]). We also noticed that the scores reported in [35] are higher than ours, however, additional datasets are needed for the perceptual loss training. Compared to BLSTM (MSE), our MetricGAN+ increases the PESQ score from 2.71 to 3.15. We also find that our model's CBAK score is lower than other state-of-the-arts; this may be because the lower threshold of the T-F mask is set to 0.05 as pointed out in Section 4.2 and hence some noise remains.

## 5. Future work

The code of MetricGAN+ is available within the SpeechBrain toolkit and we encourage the community to continuously improve its performance and training efficiency. In the following, we list some directions that are worth exploring:

1) Since MetricGAN is a black-box framework, it can be used to optimize different metrics. In [17], it was applied to optimize speech intelligibility. To the best of our knowledge, it has not been used for WER minimization under noisy conditions for a black-box ASR model (e.g., Google ASR).

2) The structure of the discriminator can be further investigated. Right now it is just a simple CNN with global average pooling. More advanced mechanisms such as attention [30, 31] may be able to replace the global pooling. In addition, if the target $Q'$ function is much more complicated (e.g., WER of an ASR model), the complexity of discriminator may also need to be increased.

3) It is time consuming to train with a replay buffer, especially when there are already lots of historical data in the buffer. Incremental learning [40] may be a good solution for this problem.

## 6. Conclusion

In this study, we proposed several techniques to improve the performance of the MetricGAN framework. We find that including noisy speech for discriminator training and applying learnable sigmoid are the most useful techniques. Our MetricGAN+ achieves state-of-the-arts results on the VoiceBank-DEMAND dataset, and the PESQ scores can increase 0.3 and 0.45 compared to MetricGAN and BLSTM (MSE), respectively. From the experimental results, we believe that the proposed framework can be further improved and applied on different tasks; therefore we made the code publicly available.